\documentclass[aps,prl,twocolumn,balance,groupedaddress,amsmath,reprint]{revtex4-1}

\usepackage[latin1]{inputenc}    % Accept european-encoded (latin1) characters.
\usepackage{graphicx}   % For eps figures
\usepackage{xspace}
\usepackage{color}
\usepackage{amsfonts}
\usepackage{changepage}
\usepackage{graphicx}
\usepackage{float}
\usepackage{amsmath}
\usepackage{amssymb}
\usepackage{tabu}
\usepackage{appendix}
\usepackage{amsmath,amssymb}
\usepackage{tikz}
\usepackage{color}

\usepackage[euler]{textgreek}
\usepackage[ugly]{units}
\usepackage{xcolor}
\usepackage{ wasysym }

\begin{document}

%Title of paper
\title{Surface acoustic wave resonators in the quantum regime} 

%Authors
\author{R.~Manenti}   
\author{M.~J.~Peterer}
\author{A.~Nersisyan}
\author{E.~B.~Magnusson}
\author{A.~Patterson}
\author{P.~J.~Leek} \email{peter.leek@physics.ox.ac.uk}

%\email{riccardo.manenti@physics.ox.ac.uk}
%Affiliation
\affiliation{Clarendon Laboratory, Department of Physics, University of Oxford, OX1 3PU, Oxford, United Kingdom}

%Date
\date{\today}

%Abstract
\begin{abstract}
We present systematic measurements of the quality factors of surface acoustic wave (SAW) resonators on ST-X quartz in the gigahertz range at a temperature of $\unit[10]{mK}$. We demonstrate a internal quality factor $Q_\mathrm{i}$ approaching $0.5$ million at $\unit[0.5]{GHz}$ and show that  $Q_\mathrm{i}\geq4.0\times10^4$ is achievable up to $\unit[4.4]{GHz}$. We show evidence for a polynomial dependence of propagation loss on frequency, as well as a weak drive power dependence of $Q_\mathrm{i}$ that saturates at low power, the latter being consistent with coupling to a bath of two-level systems. Our results indicate that SAW resonators are promising devices for integration with superconducting quantum circuits.
\end{abstract}

\maketitle

When a harmonic eigenmode of a mechanical object is well isolated from the environment and cooled to low temperature, it can approach its quantum mechanical ground state where it may be used for fundamental tests of quantum mechanics at a macroscopic scale \cite{Poot:2012}, and for applications in quantum information \cite{LaHaye:2009,O'Connell:2010,Pirkkalainen:2013} and  high-sensitivity detection \cite{Teufel:2009,Moser:2013,Tao:2014}. Research on reaching this quantum limit in mechanical resonators spans a broad range of system sizes and resonant frequencies, from meter-scale mirrors used for gravitational wave detection in the $\unit[100]{Hz}$ range \cite{Abbott:2009} to nanoscale structures that can reach resonant frequencies above $1~\rm{GHz}$ \cite{O'Connell:2010, Chan:2011,Chaste:2011}. Reaching high quality (Q) factors in such resonators is typically much harder at higher frequencies \cite{Ekinci:2005,Imboden:2007}, 
though a Q factor approaching $10^7$ has recently been demonstrated at $\unit[3.6]{GHz}$ in a silicon nanobeam \cite{Meenehan:2014}.

The integration of mechanical resonators with superconducting electric circuits is actively explored for the realization of on-chip cavity optomechanics \cite{Aspelmeyer:2014,Teufel:2011} and mechanical circuit quantum electrodynamics \cite{LaHaye:2009,O'Connell:2010,Pirkkalainen:2013}, for quantum information and detection applications in the microwave domain. 
In these investigations the mechanical resonance is typically a bulk harmonic mode of a micro-fabricated object. 
However, recent experiments have also demonstrated detection of traveling surface acoustic waves (SAWs) at the quantum level \cite{Gustafsson:2012}, as well as their interaction with a superconducting qubit \cite{Gustafsson:2014}. 
In this letter, we present a systematic study of SAW resonators operated at low temperature $T\approx\unit[10]{mK}$ at or close to the quantum regime $k_\mathrm{B}T\ll h f_{0}$ (where $f_0$ is the resonant frequency), and show that high quality factors can be reached, demonstrating their promise for integration with quantum coherent devices.
\begin{figure}[t]
\centering
\includegraphics[width=0.95\linewidth]{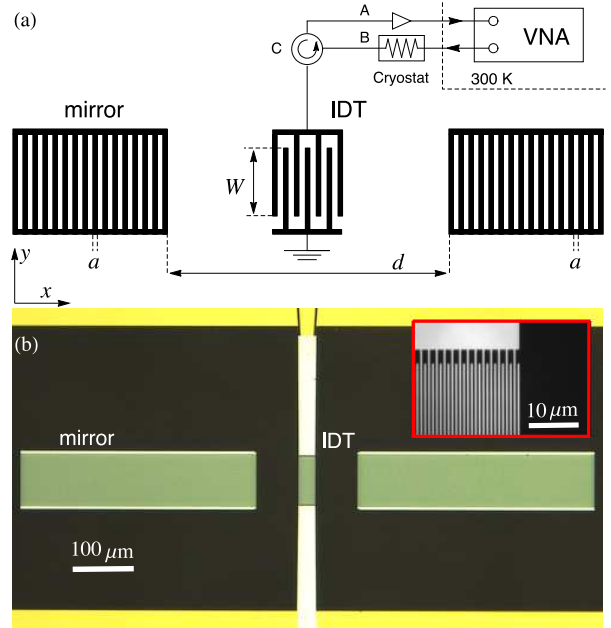}
\caption{\label{fig:1} (color online) (a) Schematic of a 1-port SAW resonator connected to the measurement setup (A: cryogenic amplifier, B: cold attenuators, C: circulator, and VNA: Vector Network Analyzer). 
 (b) Optical microscope image of a SAW resonator (device $r_3$, see Table~\ref{table:1}). Inset: magnification of the IDT electrodes of a similar device ($q_1$).}
\end{figure}

SAWs are acoustic modes of a crystal that are confined to the substrate surface and 
can be excited and detected by electric circuits on piezoelectric substrates. They have long been used in electrical engineering with a wide range of applications in communications technology \cite{Morgan:2007}. The interconversion between acoustic and electric signals is achieved using an interdigital transducer (IDT), whose periodicity matches the wavelength of the SAW at a certain frequency. A high quality reflector for a SAW can be made using an array of shorted or open circuit electrodes, or shallow grooves etched in the surface, to implement a half wavelength period modulation of the impedance of the wave medium similar to a Bragg grating in optics. A pair of such reflectors placed some distance apart on the substrate forms a Fabry-Perot cavity for SAWs \cite{Bell:1976}, the properties of which can be probed using an appropriately placed IDT (see Fig.~\ref{fig:1}). Since a SAW typically travels at a few km/s, resonators in the gigahertz range have $\mu\rm{m}$-scale wavelengths, and can thus be fabricated with standard lithographic techniques.

We investigate the quality factors of SAW resonators measured inside a dilution refrigerator at high vacuum and a temperature of $T\approx  \unit[10]{mK}$, under which conditions the dissipation of the SAW is very low. We work with the common SAW substrate ST-X quartz due to its known good performance in resonators at room temperature and weak piezoelectric coupling \cite{Morgan:2007}, which results in SAWs with little electrical character and hence good prospects for being weakly electrically coupled to environmental sources of dissipation. The reflectors and IDTs are in all cases made from superconducting aluminum such that at $10~\rm{mK}$ ohmic losses can be neglected. The resonators were characterized by measuring the complex reflection coefficient $S_{11}(f)$ at the IDT using a vector network analyzer (VNA). Using an RLC equivalent circuit model, one can derive the following expression for the reflection coefficient close to a single  resonance
\begin{equation}
S_{11}\left(f\right)=\frac{\left(Q_\mathrm{e} - Q_\mathrm{i} \right)/Q_\mathrm{e}+2iQ_\mathrm{i}\left(f-f_0\right)/f}{\left(Q_\mathrm{e} + Q_\mathrm{i} \right)/Q_\mathrm{e}+2iQ_\mathrm{i}\left(f-f_0\right)/f}.\label{eq:s11}
\end{equation}
Here $Q_\mathrm{i}$ is the internal Q-factor of the mode and $Q_\mathrm{e}$ is the external Q-factor due to the presence of the IDT and measurement port. 

\begin{figure}
\includegraphics[width=\linewidth]{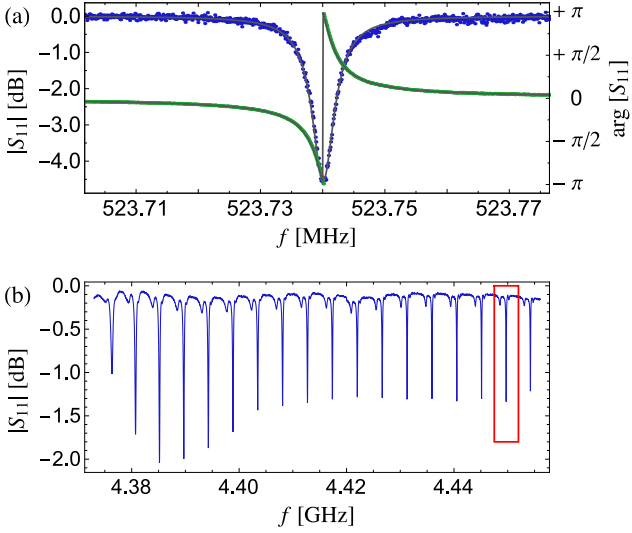}
\caption{\label{fig:2} (color online) (a) Magnitude (blue) and phase (green) of the measured reflection coefficient $S_{11}\left(f \right)$ of the SAW resonator $p_1$. Solid lines are a fit to Eq.~\eqref{eq:s11}. (b) Frequency response of device $q_7$ showing 18 supported high Q modes. The solid red box indicates the resonant mode measured in Fig. 4. A background due to the measurement setup has been subtracted in both (a) and (b).}
\end{figure}

%--------------------------------------------------------------------------------------
% Basics of SAW resonators
%--------------------------------------------------------------------------------------
The measured single-port SAW resonators can be characterized by a set of geometric parameters illustrated in Fig.~\ref{fig:1}(a) and listed in Table~\ref{table:1}. The devices have a frequency response centered at $f_0=v/\lambda_{0}$, where $v\approx\unit[3100]{m/s}$ is the SAW velocity and $a=\lambda_{0}/4$ is the electrode and space width in the lithographically defined mirrors and IDT. The mirrors are separated by an integer number of half wavelengths $d=m\lambda_0/2$ and their reflectivity is given by $R=\tanh(N_\mathrm{g}|r_\mathrm{s}|)\approx1$ in the limit $N_\mathrm{g} |r_\mathrm{s}|\gg 1$, where $N_\mathrm{g}$ is the number of electrodes in the mirror and $r_\mathrm{s}$ is the reflectivity of each electrode. In this limit, the mirrors have a high reflectivity within their first stopband $\Delta f_{\mathrm{1SB}}=2f_0 |r_s|/\pi$. The IDT excites and detects SAWs in a broader frequency range than the mirrors (due to being smaller spatially) and has a bandwidth $\Delta f_{\mathrm{IDT}}=1.8f_0 /N_\mathrm{t}$, where $N_\mathrm{t}$ is the number of electrodes in the IDT \cite{Morgan:2007}. Thus within $|f-f_0|<\Delta f_{\mathrm{ 1SB}}/2$ the resonator supports high quality resonant modes, measurable via the IDT. Since the electrode reflectivity is small ($|r_\mathrm{s}|\approx 0.2\%$ for our devices), the resonant modes partly penetrate into the mirrors to a penetration depth $L_\mathrm{p}=a/|r_\mathrm{s}|$ \cite{Morgan:2007}. The devices therefore behave as acoustic Fabry-Perot cavities, with cavity length $L_\mathrm{c}=d+2L_\mathrm{p}$ and free spectral range $\mathrm{FSR}=v/L_\mathrm{c}=2f_0/(d/2a+1/|r_\mathrm{s}|)$. When $\mathrm{FSR}>\Delta f_{\rm 1SB}$, a single mode resonance is observed within the mirror stopband, whereas for longer resonators, for which $\mathrm{FSR}<\Delta f_{\rm 1SB}$, multiple resonances are observed (see Fig.~\ref{fig:2}).

\begin{figure}[t]
\begin{center}
\includegraphics[width=\linewidth]{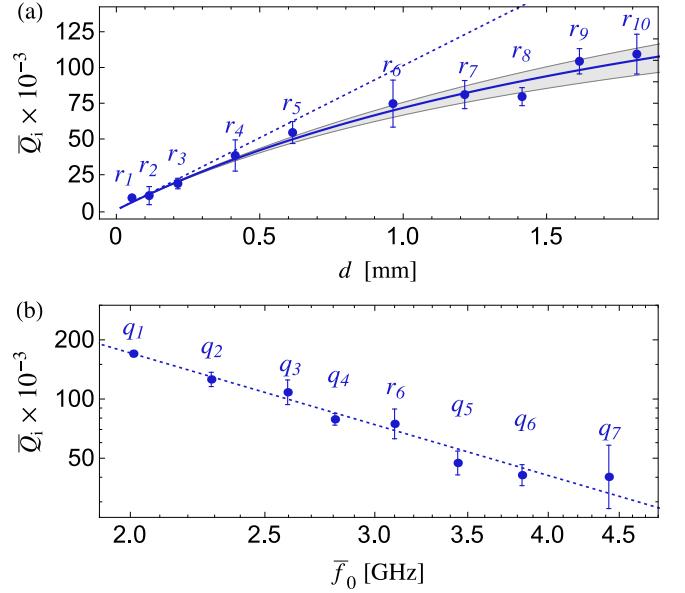}
\caption{\label{fig:3} (color online) (a) $\overline{Q}_\mathrm{i}$ versus $d$ for devices $r_1-r_{10}$. Solid line is a fit to Eq.~\eqref{eq:3}. Shaded area indicates one standard deviation of $\alpha_p$ and the dashed line is the linear part of this fit ($\alpha_p=0$).
(b) $\overline{Q}_\mathrm{i}$ versus $\overline{f}_0$ for devices $q_1-q_7$ and $r_6$. 
Dashed line is a fit to $\overline{Q}_\mathrm{i}=c_1 f^{-c_2}$.}
\end{center}
\end{figure}

%--------------------------------------------------------------------------------------
% Preliminary investigation with 500 MHz devices
%--------------------------------------------------------------------------------------
We have initially performed a comprehensive study of SAW resonators at a wavelength of $\lambda_0=\unit[6]{\mu m}$ ($f_0=\unit[524]{MHz}$) to determine how $Q_\mathrm{e}$ and $Q_\mathrm{i}$ depend on transducer and grating geometry. This frequency was chosen for compatibility with standard photolithography for which the feature sizes of $a=\unit[1.5]{\mu m}$ are achievable and a large number of devices could be fabricated on a single wafer. 
In this initial investigation we determined that, in accordance with SAW theory  \cite{Aref:2015}, the external quality factor follows $Q_\mathrm{e} \propto L_\mathrm{c}/N_\mathrm{t}^2$ and the internal quality factors were limited by grating reflectivity, following
\begin{equation}
Q_\mathrm{g}=\frac{\pi \left(d+2L_\mathrm{p}\right)}{\lambda_0\left[1-\tanh \left(|r_\mathrm{s}|N_\mathrm{g}\right)\right]}. \label{eq:2}
\end{equation}
Based on these observations, we designed a device to realise a high $Q_\mathrm{i}$ (within the confines of our chip geometry) with widely spaced long gratings (device $p_1$ see Table~\ref{table:1}). This device exhibits $Q_\mathrm{e}=1.16\times10^5$, $Q_\mathrm{i}=4.53\times10^5$ and the frequency response is shown in Fig.~\ref{fig:2}(a). Note that this measurement is not in the quantum regime, since $k_\mathrm{B}T\approx hf_0$.

%--------------------------------------------------------------------------------------
% Cavity length dependence
%--------------------------------------------------------------------------------------
 We next proceeded to fabricate SAW resonators at higher frequencies, using electron beam lithography. In Fig.~\ref{fig:3}(a) we plot $Q_\mathrm{i}$ for a series of resonators with $\lambda_0=1.0~\rm{\mu m}$ ($f_0\approx\unit[3.1]{GHz}$), for which the distance between the two gratings $d$ was varied over the range $\unit[0.05-1.8]{mm}$ with all other geometric parameters fixed (see Table~\ref{table:1}). The number of modes seen increases from 1 for device $r_1$ to 65 for device $r_{10}$. For devices with more than 5 modes ($r_5-r_{10}$), we took the average $Q_\mathrm{i}$ for the 5 modes at the center of the grating stopband where the reflectivity is highest.
We expect $Q_\mathrm{i}$ to be dominated by the grating reflectivity at low $d$ and saturate at high $d$ due to propagation losses. 
The data follow this trend and they can be fitted with the following relation
\begin{equation}
Q_{\mathrm{i}}=\left(\frac{1}{Q_\mathrm{g}}+\frac{v\, \alpha_{\mathrm{p}}}{\pi f_0}\right)^{-1}. \label{eq:3}
\end{equation}
From a fit to Eq.~\eqref{eq:3}, we determine the electrode reflectivity to be $\left|r_\mathrm{s}\right|=0.002$, and we can place an upper limit on the propagation loss at $\unit[3.1]{GHz}$ of $\left.\alpha_{\mathrm{p}}\right|_{\unit[10]{mK}}<\unit[0.015]{mm^{-1}}$ corresponding to a phonon mean free path of $l=1/\alpha_{\mathrm{p}} \approx \unit[6]{cm}$.

\begin{figure}[t!]
\begin{center}
\includegraphics[width=\linewidth]{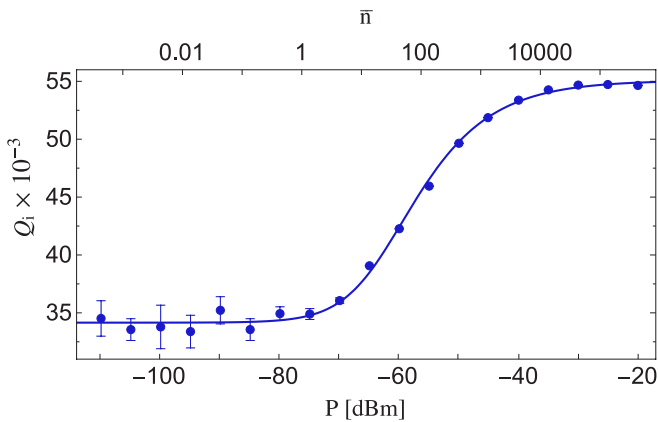}
\caption{ (color online) $Q_\mathrm{i}$ versus drive power $P$
for the resonant mode $f_0=\unit[4.449]{GHz}$ of device $q_7$. Solid line: fit to $Q_\mathrm{i}=\left(v\,\alpha_{_{\mathrm{TLS}}}/\left(\pi f_0 \right)  +1/Q_\mathrm{rl}\right)^{-1}$; $\overline{n}$ is the mean number of phonons occupying the resonator due to the coherent drive $P$.
\label{fig:5}}
\end{center}
\end{figure}

%--------------------------------------------------------------------------------------
% Frequency dependence
%--------------------------------------------------------------------------------------
We then moved on to investigate the frequency dependence of $Q_\mathrm{i}$, using a device geometry with large $d=1929\lambda_0/2$, over the range $\unit[2.0-4.4]{GHz}$ (devices $q_1-q_7$, see Table~\ref{table:1}).
These long devices all display around 20 resonant modes within the stopband of the reflectors (see Fig.~\ref{fig:2}(b) for the frequency response of device $q_7$). In Fig.~\ref{fig:3}(b) we plot the dependence of the average internal quality factor $\overline{Q}_\mathrm{i}$ of all modes of each device against average mode frequency $\bar{f}_0$. Error bars represent the standard deviation of $Q_\mathrm{i}$ from the set of resonant modes. 
The quality factor is seen to decrease with frequency, and the linear trend observed on a log-log scale indicates a polynomial dependence. We fit the data to $\bar{Q}_\mathrm{i}=\pi f/v \alpha_{\mathrm{p}}=c_1 f^{-c_2}$ and find $c_1=\unit[719\pm85]{} $, $c_2=2.07\pm 0.13$. This strongly suggests $\alpha_\mathrm{p}\propto f^3$, agreeing with a model developed for SAW propagation loss in the high frequency and low temperature limit \cite{Sakuma:1974}.

%--------------------------------------------------------------------------------------
% Power dependence
%--------------------------------------------------------------------------------------
We conclude our investigation with a measurement of the internal quality factor of one of the modes of device $q_7$ ($f_0=\unit[4.449]{GHz}$) at low drive powers such that the average resonator phonon population reaches the regime $\overline{n} \! \ll \!1$. Figure~\ref{fig:5} shows the dependence of $Q_\mathrm{i}$ on drive power $P$. In this case, the VNA was connected to device $q_7$ through highly attenuated microwave lines with an overall attenuation from the instrument to the sample of $-\unit[67\pm1]{dB}$. This allows us to calculate the average phonon population $\overline{n}$ resulting from the coherent drive, shown in the upper scale of Fig.~\ref{fig:5}. 
A clear reduction in $Q_\mathrm{i}$ is observed as $P$ is reduced, saturating at low power at a value of $Q_{\mathrm{i}0}=34500$. 
A similar dependence has been observed in bulk mechanical resonators \cite{Goryachev:2012,Galliou:2013} and electromagnetic coplanar waveguide resonators (CPWR) \cite{Macha:2009} at low temperature, and has been attributed to coupling to a bath of two level systems (TLS). The analytical expression of the loss rate caused by a TLS bath is given by \cite{Phillips:1987}
\begin{equation*}
\alpha_{_{\mathrm{TLS}}} = \frac{ 2 \pi^2 f_0 n_0 \gamma^2}{\rho v^2} \left( 1+\frac{P}{P_\mathrm{c}}\right)^{-0.5}\tanh \left(\frac{h f_0}{2k_\mathrm{B}T}\right)\!, 
\end{equation*}
where $n_0$  is the density of states of the TLS, $\rho$ is the density of the crystal, $P_\mathrm{c}$ is a critical power and $\gamma$ describes the strength of the coupling between the TLS and the phonons. The internal quality factor is related to $\alpha_{\mathrm{TLS}}$ by $
Q_\mathrm{i}=\left(v\,\alpha_{_{\mathrm{TLS}}}/\pi f_0  +1/Q_\mathrm{rl}\right)^{-1},
$
where $Q_\mathrm{rl}$ takes into account remaining losses. We find that our data fits well to this expression with $P_{\mathrm{c}}=\unit[-65.7]{dBm}$ and $n_0 \gamma^2=\unit[4.5\times10^{4}]{J/m^3}$ (three orders of magnitude lower than in glasses \cite{Golding:1976}). The ratio $Q_{\mathrm{i}0}/Q_\mathrm{rl}\approx0.6$ is much higher than typically seen in CPWRs \cite{Megrant:2012}, indicating that any TLS contribution to the loss is considerably less in the SAW case. Such a difference is qualitatively in agreement with an electric field coupling to the TLS bath, since in a weak piezoelectric such as quartz, only a small fraction of the SAW energy is electrical.
%--------------------------------------------------------------------------------------
% Prospects
%--------------------------------------------------------------------------------------

We finally comment on the prospects for realizing strong coupling cavity QED between a SAW resonator and a superconducting qubit, which requires a coupling strength $g$ between qubit and resonator exceeding the linewidths of both \cite{Haroche:2006}. The linewidth for resonator $q_7$ in the quantum regime is $\kappa=f/Q_\mathrm{i}\approx \unit[130]{kHz}$, while a superconducting qubit on quartz may be expected to have linewidth $\gamma\lesssim 1~\rm{MHz}$ from experiments on GaAs, a similar piezoelectric substrate \cite{Gustafsson:2012}. The coupling strength can be estimated as $\hbar g=e\beta V^{\mathrm{rms}}_0$  \cite{Blais:2004} where the electric potential due to the vacuum fluctuations of the SAW mode $V^{\mathrm{rms}}_0\approx 20~\rm{nV}$ for a geometry similar to device $q_7$, and $\beta$ is a dimensionless parameter that takes into account the geometric match of the qubit to the SAW. For a superconducting qubit with well chosen geometry, $g\approx(0.2)eV^{\mathrm{rms}}_0/\hbar\approx 2\pi\times1~\rm{MHz}$ should be easily achievable. Coupling strengths in the $\unit[10-100]{MHz}$ range could be achieved in stronger piezoelectrics such as $\mathrm{LiNbO}_3$ \cite{Morgan:2007} or ZnO \cite{Magnusson:2015}. It should also be possible to couple in a similar way a wide variety of other solid state quantum systems, such as quantum dots and crystal defect centre spins, to SAW devices \cite{Schuetz:2015}.

%--------------------------------------------------------------------------------------
% Summary
%--------------------------------------------------------------------------------------
In summary, we have fabricated SAW resonators with a range of geometries and frequencies in the gigahertz range and measured them at cryogenic temperatures, demonstrating quality factors up to $4.5\times10^5$. By measuring a range of different resonator lengths, we are able to place an upper limit on propagation loss at $3.1~\rm{GHz}$ and $10~\rm{mK}$ of $\alpha_\mathrm{p}\leq 0.015~\rm{mm}^{-1}$. We observed a frequency dependence of $Q_\mathrm{i}$ for long resonators that suggests $\alpha_\mathrm{p}\propto f^3$. In a highly isolated measurement setup, we observed a clear power dependence of the quality factor of a $4.4~\rm{GHz}$ resonator consistent with coupling to a two-level system bath. We have demonstrated internal quality factors in the $10^4$ range up to $4.4~\rm{GHz}$, which should provide motivation for experiments that integrate SAW resonators with quantum coherent devices such as superconducting qubits.

We would like to thank D. Hewitt for technical contributions to the project and P. Delsing, T. Aref, M. K. Ekstr\"{o}m and S. H. Simon for fruitful discussions. This work has received funding from the UK Engineering and Physical Sciences Research Council under Grant Nos. EP/J001821/1 and EP/J013501/1, and the People Programme (Marie Curie Actions) of the European Union's Seventh Framework Programme (FP7/2007-2013) under REA Grant Agreement No. [304029].

\begin{table}[t]
\caption{\label{table:1} Parameters of the measured SAW resonators. For device $p_1$, the thickness of
the aluminum layer is $h=\unit[100]{nm}$ and the geometric parameters are $W=600a$, $N_t=51$ and $N_g=1500$; for all other devices $h=\unit[30]{nm}$, $W=400a$, $N_t=71$ and $N_g=1000$. }
\begin{tabular}{c r c r r r c   }
\hline\hline
 & & \\[-8pt]
     Device   & $\unit[a]{[nm]}$ &$\unit[\overline{f}_0]{[GHz]}$ &$ d/2a $&$\overline{Q}_\mathrm{e}/10^3$& $\overline{Q}_\mathrm{i}/10^3$&$\overline{Q}_\mathrm{i}\overline{f}_0/10^{14}$\\[2pt]
\hline
 & & \\[-6pt]
$p_1$ & $\unit[1500]{}$ &$\unit[0.52]{}$  & $ 1051 $ &116&$453$ & $2.36$\\[3pt]
\hline \\[-9pt]
$r_1$ & $\unit[250]{}$ &$\unit[3.11]{}$ & $ \phantom{0}109 $  &24&$\phantom{00}8.8$&$0.27$\\[3pt]
$r_2$ & $\unit[250]{}$ &$\unit[3.12]{}$    & $ \phantom{0}229 $  &18&$\phantom{0}10.4$&$0.32$\\[3pt]
$r_3$ & $\unit[250]{}$ &$\unit[3.11]{}$  & $ \phantom{0}429 $ &98&$\phantom{0}18.8$&$0.62$\\[3pt]
$r_4$ & $\unit[250]{}$ &$\unit[3.11]{}$   & $ \phantom{0}829 $ &167&$\phantom{0}38.4$&$1.19$\\[3pt]
$r_5$ & $\unit[250]{}$ &$\unit[3.10]{}$                   & $ 1229 $ &363&$\phantom{0}54.5$&$1.74$\\[3pt]
$r_6$ & $\unit[250]{}$ &$\unit[3.09]{}$                   & $ 1929 $ &657&$\phantom{0}74.7$&$2.32$\\[3pt]
$r_7$ & $\unit[250]{}$ &$\unit[3.09]{}$              & $ 2429 $ &473&$\phantom{0}81.0$&$2.52$\\[3pt]
$r_8$ & $\unit[250]{}$ &$\unit[3.10]{}$         & $ 2829 $ &843&$\phantom{0}79.6$&$2.45$\\[3pt]
$r_9$ & $\unit[250]{}$ &$\unit[3.11]{}$         & $ 3229 $ &1230&$\phantom{0}103$&$3.18$\\[3pt]
$r_{10}$ & $\unit[250]{}$ &$\unit[3.08]{}$       &   $ 3629 $ &927&$\phantom{0}109$&$3.23$\\[3pt]
 \hline \\[-9pt]
$q_1$ & $\unit[390]{}$ &$\unit[2.01]{}$  & $ 1929 $  &242&$171$ &$3.43$\\[3pt]
$q_2$ & $\unit[340]{}$ &$\unit[2.29]{}$  &  $ 1929 $ &499&$126$ &$2.88$\\[3pt]
$q_3$ & $\unit[300]{}$ &$\unit[2.60]{}$  & $ 1929 $ &174&$108$ &$2.81$\\[3pt]
$q_4$ & $\unit[275]{}$ &$\unit[2.81]{}$  & $ 1929 $ &232&$\phantom{0}78.8$ &$2.21$\\[3pt]
$q_5$ & $\unit[225]{}$ &$\unit[3.44]{}$  & $ 1929 $ &445&$\phantom{0}47.3$ &$1.63$\\[3pt]
$q_6$ & $\unit[200]{}$ &$\unit[3.83]{}$  & $ 1929 $ &358&$\phantom{0}41.0$ &$1.57$\\[3pt]
$q_7$ & $\unit[175]{}$ &$\unit[4.42]{}$ & $ 1929 $ &528&$\phantom{0}40.2$ &$1.78$\\[3pt]
 & & \\[-10pt]
\hline\hline\\
\end{tabular}
\end{table}

\bibliographystyle{apsrev4-1}
\bibliography{Manenti150521} 

\clearpage

\end{document}